\documentclass[aip,jcp,reprint,groupedaddress]{revtex4-1}

\usepackage{amsmath}
\usepackage{tabularx}
\newcolumntype{Y}{>{\centering\arraybackslash}X}
\usepackage{graphicx}  
\usepackage{dcolumn}   
\usepackage{amssymb}   

\begin{document}

\title{Non-Equilibrium Surface Tension of the Vapour-Liquid Interface of Active Lennard-Jones Particles}

\author{Siddharth Paliwal}
\email{s.paliwal@uu.nl}
\author{\normalfont\textsuperscript{b)}}
\author{Vasileios Prymidis}
\altaffiliation{These authors contributed equally to this work}
\author{Laura Filion}
\author{Marjolein Dijkstra}
\email{m.dijkstra@uu.nl}

\affiliation{Soft Condensed Matter, Debye Institute for Nanomaterials Science, Utrecht University, Princetonplein 5, 3584 CC Utrecht, The Netherlands}

\date{\today}

\begin{abstract}
We study a three-dimensional system of self-propelled Brownian particles interacting via the Lennard-Jones potential. Using Brownian Dynamics simulations in an elongated simulation box, we investigate the steady states of vapour-liquid phase coexistence of active Lennard-Jones particles  with  planar interfaces. We measure the normal and tangential components of the pressure tensor along the direction perpendicular to the interface and verify mechanical equilibrium of the two coexisting phases. In addition, we determine the non-equilibrium interfacial tension by integrating the difference of the normal and tangential component of the pressure tensor, and show that the surface tension as a function of strength of particle attractions is well-fitted by simple power laws. 
Finally, we measure the interfacial stiffness using capillary wave theory and the equipartition theorem, and find a simple linear relation between surface tension and  interfacial stiffness with a proportionality constant characterized by an effective temperature.  
\end{abstract}


\maketitle

\section{Introduction}
Many-particle systems that are driven out-of-equilibrium exhibit intriguing collective behaviour like clustering, laning, swarming, but also phenomena that resemble equilibrium phase behavior such as crystallization, condensation and phase separation. As a consequence, there has been considerable interest in exploring the applicability of  equilibrium statistical physics concepts, such as pressure, chemical potential, and surface tension, to describe  non-equilibrium steady states resembling phase coexistence.\cite{Narayan2007,Wittkowski2014,  Takatori2015, Marconi2015,  Solon2015, Falasco2015,Speck2015,Ginot2015a,Speck2016,herminghaus2016phase,clewett2016minimization,clewett2012emergent, Wysocki2014,Stenhammar2014,Tailleur2008a,Fily2012,Bialke2013a,Lee2016,Farage2015,Redner2013,Redner2013a,Cates2013,Stenhammar2013, Stenhammar2014,Fily2014,Takatori2014,zottl2014hydrodynamics,Bialke2015,Takatori2015b,Yan2015} 

Very recently, it was shown by experiments and simulations that steady states of a granular gas under vibration resemble phase coexistence of  a dilute gas and a dense liquid phase that follows the lever rule, whereas the coexisting densities are well-predicted by a Maxwell equal-area construction to the   equation of state. \cite{herminghaus2016phase,clewett2016minimization}    Additionally, it was observed that these granular gases   form patterns  that resemble   spinodal decomposition with a coarsening dynamics that proceeds via  the same spatio-temporal scaling laws as in equilibrium molecular fluids. \cite{herminghaus2016phase,clewett2016minimization}  In the case of molecular fluids, the coarsening is driven by a reduction of  the interfacial  area and thereby a minimization of the interfacial energy. For granular gases, it was  found that the coarsening dynamics can indeed be explained by the emergence of  a positive non-equilibrium surface tension that is predominately determined by the anisotropy in the kinetic energy part of the stress tensor in contrast to  the surface tension in molecular fluids, which is mainly determined by energetic interactions.\cite{clewett2012emergent} 

Another model system that has recently received huge interest  is a system of active Brownian particles suspended in a solvent that incessantly convert energy from the local environment into directed motion, and are thus inherently out-of-equilibrium. The self-propulsion can be generated through a variety of mechanisms, for example,
by hydrodynamic flows around a bacterium \cite{Berg2004,Czirok1996,Aranson2013}, self-diffusiophoresis\cite{palacci2010sedimentation}, bubble propulsion \cite{Gao2012,Solovev2009}, local demixing of a near critical solvent mixture \cite{buttinoni2013dynamical,volpe2011microswimmers},  thermophoresis \cite{Golestanian2012}, Marangoni flows \cite{Hanczyc2007,Thutupalli2011},   self-electrophoresis \cite{Dong2016},  etc.
In the simplest model of active Brownian particles, the particles perform directed motion with a constant self-propulsion speed, whereas the Brownian motion is described by stochastic translational forces as well as stochastic rotational forces that alter the direction of the persistent motion. 
In the case where these particles interact with purely repulsive interactions, dense clusters of particles in a dilute phase were observed at sufficiently high self-propulsion speeds in numerical simulations and in theory, a phenomenon termed as motility-induced phase separation (MIPS).\cite{Tailleur2008a,Fily2012,Bialke2013a,Redner2013,Redner2013a,Cates2013,Stenhammar2013, Stenhammar2014,Fily2014,Wysocki2014,Takatori2014,zottl2014hydrodynamics,Farage2015,Lee2016}
Using large system sizes and an elongated simulation box, a stable phase separation between a dense and dilute phase separated by  planar interfaces was also achieved.\cite{Bialke2015} Remarkably,  the mechanical interfacial tension as determined by integrating the anisotropy of the pressure tensor in these simulations turns out to be negative. In the case of a negative surface tension in equilibrium fluids, the system can lower its free energy by creating more interface, and hence the phase separation is unstable. This intuitive interpretation of a negative surface tension is thus at odds with the observation of a stable motility-induced phase separation, thereby questioning the mechanical definition of surface tension and its equilibrium-like interpretation.

On the other hand, phase separation has also been observed in systems of self-propelled
particles  interacting with attractive interactions. \cite{Schwarz2012,Redner2013,Redner2013a,Mognetti2013,Farage2015,Prymidis2015b,Prymidis2016} Interestingly, a reentrant phase behavior was found in simulations of active colloidal particles interacting via  Lennard-Jones interactions. \cite{Redner2013,Redner2013a}  Phase-separated states were observed at low as well as high activities, and  homogeneous states were found at intermediate activities. \cite{Redner2013,Redner2013a} At high activity, the phase separation resembles the motility-induced phase separation as observed for active repulsive particles, whereas for low activity the phase separation is caused by the attractive particle interactions and a kinetically arrested attractive gel phase is observed reminiscent of spinodal decomposition.  \cite{Redner2013,Redner2013a}  However,  it is yet unknown whether the coarsening dynamics  of the spinodal structure of active Brownian Lennard-Jones particles  bears any similarities with that of molecular fluids. 

We thus conclude that many out-of-equilibrium steady states show behavior reminiscent to that observed for equilibrium fluids such as condensation, crystallization, and phase separation. More surprisingly, also the kinetics of the phase separation  displays striking similarities with the equilibrium counterparts. Vibrated granular systems exhibit spinodal decomposition with a coarsening dynamics that emerges from the presence of a non-equilibrium positive interfacial tension, whereas  active repulsive particles show motility-induced phase separation with a negative surface tension implying that work is released by creating more interface while keeping the volume of the system fixed. \cite{Bialke2015}  In order to gain more insight in the concept of an interfacial tension in non-equilibrium systems, we investigate the interfacial tension and  stiffness of a vapour-liquid interface of active Lennard-Jones systems.  Many reasons justify this choice of  system. First of all, the bulk and interfacial behavior as well as the critical behavior of passive Lennard-Jones systems have been extensively studied over the past decades and we are provided with a wealth of information on the equilibrium passive counterpart of this model.  \cite{smit1992phase,Vrabec2006,Watanabe2012, noro2000extended, dunikov2001corresponding}  Secondly, the computational efficiency of the model makes it  very convenient and attractive  for computer simulations. Furthermore, and perhaps more importantly, the system undergoes a vapour-liquid phase transition due to particle attractions for very low but also  high activities of the particles, which correspond to both quasi-equilibrium and fairly out-of-equilibrium regimes. It is thus an ideal system to study systematically  the effect of self-propulsion on the properties of the phase transition and of the interface   as one can slowly switch on the activity of the system and drive the system  further out-of-equilibrium, contrary to the case of motility-induced phase separation.

To this end, we study   a stable vapour-liquid phase coexistence of  isotropic  self-propelled Brownian particles interacting with a truncated and shifted Lennard-Jones potential using Brownian dynamics simulations. Here, our investigation builds upon our previous work, in which we determined the vapour-liquid binodals as a function of rotational diffusion rate and self-propulsion speed of active Lennard-Jones particles. \cite{Prymidis2016}
We use the overdamped Langevin equation to simulate the dynamics of
the particles considering an implicit solvent. In order to stabilize direct coexistence, we employ an elongated simulation box, in which the planar interfaces align with the shortest dimensions of the box.  
We measure the normal and tangential components of the pressure tensor in the direction perpendicular to the interface by  employing a local expression for the pressure tensor in active systems.  \cite{Winkler2015,Solon2014,Solon2015,Dijkstra2016a}
The non-equilibrium interfacial tension is measured by integrating the difference of the normal and tangential component of this  pressure tensor. \cite{Evans1979} We calculate the non-equilibrium surface tension  for different
combinations of self-propulsion speed and rotational diffusion
rate, and demonstrate that the trends of the surface tension can be fitted by simple power laws. In addition, we also apply capillary wave theory to understand the non-equilibrium relationship of interfacial tension and stiffness coefficient.

This paper is organized as follows. 
In section \ref{sec:methods}, we describe our model and the dynamics used in our numerical study. In section \ref{sec:pressure}, we present our method which we used to measure the pressure tensor profiles and surface tension. 
We then discuss the density and pressure profiles in Sections \ref{sec:rho_prof} and \ref{sec:pressure_prof}, 
respectively.
We determine the non-equilibrium interfacial tension in Section \ref{sec:surf_tension}, and show that the surface tension as a function of strength of particle attractions is well-fitted by a simple power law. 
Finally, we present our results on the interfacial stiffness as obtained from the application of capillary wave theory and  equipartition theorem in Section \ref{sec:capillary} and discuss its relation to the  surface tension measured in Section  \ref{sec:surf_tension}. We end with some conclusions in Section \ref{sec:conclusions}.

 
\section{Model and Methods}
\label{sec:methods}
 We consider a three-dimensional system consisting of self-propelled spherical particles suspended in a solvent. The particles interact via a truncated and shifted Lennard-Jones potential $U(r_{ij})$ given by 
	\begin{align}
		U(r_{ij}) &= U_{LJ}(r_{ij}) - U_{LJ}(r_{cut})  && r \leq r_{cut}	\nonumber \\
						&= 0 	&& r > r_{cut}	\nonumber 
	\end{align}
	with 
	\begin{align}
		U_{LJ}(r_{ij}) &= 4\epsilon\left[
		\left(\frac{\sigma}{r_{ij}}\right)^{12} - \left(\frac{\sigma}{r_{ij}}\right)^6
		\right], 						
	\end{align}
where $r_{ij} = |\mathbf{r}_j - \mathbf{r}_i|$ is the center-of-mass distance, $\mathbf{r}_i$ the position of particle $i$, $\sigma$ the particle length scale, and 
$\epsilon$  the strength of the particle interaction. We set the cut-off radius $r_{cut}=2.5\sigma$ in all our simulations.  In addition, we associate a three-dimensional unit vector $\mathbf{u}_i$ to particle $i$  that indicates the direction of the self-propelling force. 

To describe the translational and rotational motion of the individual colloidal particles inside the solvent we use the overdamped Langevin equations. 
\begin{align}
	\frac{d \mathbf{r}_i}{dt} &= -\frac{1}{\eta}
	\sum_{j \neq i}^{} \frac{\partial U (r_{ij})}{\partial \mathbf{r}_i}
	+ v_0\mathbf{u}_i
	+ \sqrt{2 D_{t}} \boldsymbol{\Lambda}_i^{t},
	\label{eqn:trans}
	\\
	\frac{d\mathbf{u}_i}{dt} &= \sqrt{2D_r}\left(\mathbf{u}_i \times \boldsymbol{\Lambda}_i^r \right), 
	\label{eqn:rotation}
\end{align}
where  $D_{t}$ is the translational diffusion coefficient given by the Stokes-Einstein relation $D_{t} = 1/(\beta_s \eta$), 
$\eta$ is the damping coefficient due to drag forces from the solvent,  $\beta_s = 1/k_BT_s$ is the inverse temperature of the solvent bath with $k_B$ the Boltzmann constant, and $T_s$  the bath temperature. $D_r$ is the rotational diffusion coefficient and $v_0$ denotes the self-propulsion speed. The vectors $\boldsymbol{\Lambda}_i^{t}$ and $\boldsymbol{\Lambda}_i^{r}$ are unit-variance Gaussian distributed random vectors with zero mean and variation 
	\begin{align}
	\left\langle \boldsymbol{\Lambda}_i^{t,r}(t) \right\rangle  &= 0,
	\\
	\left\langle \boldsymbol{\Lambda}_i^{t,r}(t) \boldsymbol{\Lambda}_j^{t,r}(t') \right\rangle
		&= \mathbb{I}_3 \delta_{ij} \delta(t-t')
	\end{align}
where $\mathbb{I}_3$ is the $3\times3$ identity matrix. The angular brackets $\langle\dots\rangle$ denote an average over different realizations of the noise. We also normalize the unit vector $\mathbf{u}_i$ of each particle $i$, after each iteration of Eq.~\ref{eqn:rotation} in order to prevent a drift with time.

We perform Brownian dynamics simulations in an elongated box  with dimensions $L\times L\times 6L$ for all  cases except in Section~\ref{sec:capillary}. The elongated shape of the simulation box stabilizes a phase coexistence with a planar interface between the two phases in the simulations. We apply periodic boundary conditions in all three directions and fix our $z$-coordinate axis along the longest edge of the box. The number of particles in our simulations is approximately $N=2500$ and the density of the system is kept fixed for all simulations at $\rho\sigma^3=0.1333$. We numerically integrate the equations of motion, Eq. (\ref{eqn:trans}) and Eq. (\ref{eqn:rotation}), using the  Euler-Maruyama scheme. \cite{Higham2001}
We set $\sigma$, $1/\beta_s$ and $\tau=\sigma^2 /D_{t}$ as our units of length, energy, and time, respectively. We use a time step $dt=2\times10^{-5}\tau$ for the numeric integration of the equations of motion. 
In equilibrium, $\epsilon$ is inversely proportional to $k_BT_s$ and either parameter could be varied to control the temperature. Here we keep the temperature of the bath fixed by keeping $\beta_s$ constant and vary $\epsilon$ to mimic the change in temperature of the colloidal particles. We employ the dimensionless temperature $T=k_BT_s/\epsilon$  following Ref.~\onlinecite{Prymidis2016}. In addition, we define the P\'eclet number  as $\textrm{Pe}=v_0/\sigma D_r$ which parameterizes   the persistence length of the active motion. We investigate the interfacial properties of the system in the P\'eclet number range of $0-8$. To change the P\'eclet number, we either increase the self-propulsion speed at  fixed  rate of rotational diffusion coefficient ($D_r\tau=20$), or we decrease the rotational diffusion coefficient at fixed  self-propulsion	speed ($\upsilon_0\tau\sigma^{-1}=8$).
The choice of parameters studied here is exactly the same as in Ref. 45 where the vapor-liquid binodals have been mapped out. Note that in Ref. 44, it was shown that a percolating network state could separate the fluid from the vapour-liquid coexistence region when the system was sufficiently far from equilibrium. However, for the parameters studied here, as argued in Ref. 45, there are no signatures of a percolating state within the coexistence regions.

For each set of simulation parameters, we let the systems reach a steady state by running the simulations for $\approx600\tau$ and then collect data	for another $1200\tau$ by measuring the quantities of interest every $100$ time steps. We also fix the center of mass of the system at the origin of the $z$-axis in order to prevent the drift of the liquid slab that coexists with the gas by regularly shifting the coordinates of the particles at fixed time intervals.


	\begin{figure}
	\centering
	\includegraphics[width=\columnwidth]{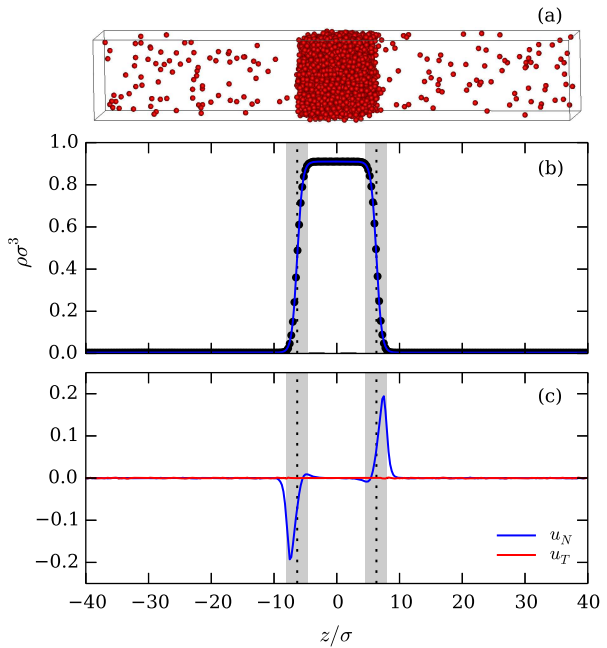}
	\caption{(a) A typical snapshot of the simulation box showing a steady state of 
		a vapour-liquid phase coexistence of active Lennard-Jones particles with a self-propulsion speed  $v_0\tau\sigma^{-1}=28$ at a temperature $k_BT_s/\epsilon=0.2$, and rotational diffusion rate $D_r\tau=20$.
		The dense liquid slab is in the middle of the box and is separated from the vapour phase by two planar interfaces.
		(b) Local density profile $\rho(z)$. Data points correspond to simulation measurements and the continuous line is the fit using Eq. (\ref{eqn:tanh_rho}).
		 Dotted lines indicate the location $z_0$ of the interface according to Eq. (\ref{eqn:tanh_rho}) and the shaded areas denote  the interfacial regions ($z_0-D/2,z_0+D/2$).
		(c) Profiles of the components of the orientation vector $\mathbf{u}(z)=\left\langle\mathbf{m}(z)\right\rangle/\left\langle\rho(z)\right\rangle$
		($u_N=u_z$ and $u_T=(u_x+u_y)/2$).
		}
	\label{fig:fig1}
	\end{figure}

\section{Pressure tensor}
\label{sec:pressure}
The concept of a non-equilibrium pressure in active systems has received a lot of attention in recent years. Various approaches have been followed to derive a microscopic expression for the bulk pressure of an active-particle  system.  It has already been shown that an extra swim pressure contribution arises due to the self-propulsion of the particles in active matter using a micromechanical, \cite{Takatori2014} a virial, \cite{Winkler2015,Speck2015,Speck2016,Lee2016} or a stochastic thermodynamics  formulation.  \cite{Speck2016} Solon and co-workers have argued that in the case of an active system interacting with anisotropic interactions the pressure depends on the wall-particle interactions, which implies that pressure is not a state function. \cite{Solon2014} In the case of isotropic interactions, the various approaches yield consistent results for the microscopic expression of the bulk pressure. Furthermore, a microscopic definition for the local stress tensor has been derived from the stationary probability distribution function by using the Fokker-Planck equation.  \cite{Marconi2015,Marconi2016b,Falasco2015,Solon2014,Solon2015,Bialke2015,Dijkstra2016a,Rodenburg2016b}

In order to simulate direct coexistence between an active vapour and liquid phase, we employ an elongated simulation box with the long axis along the $z$-direction and in which the two coexisting phases are separated by interfaces parallel to the $xy$-plane. Hence, the system is only inhomogeneous in the $z$-direction and consequently, the pressure tensor contains only two independent components, the normal component along the direction perpendicular to the interfaces, $P_{N}(z)=P_{zz}(z)$, and the transverse component,  $P_{T}(z)=(P_{xx}(z)+P_{yy}(z))/2$, which is the average of the $xx-$ and $yy-$components due to the symmetry of the system in the $xy$-plane. The non-diagonal components of the pressure tensor vanish due to hydrostatic equilibrium, which we   verified  in our simulations.

As described in the references \onlinecite{Solon2015,Dijkstra2016a,Rodenburg2016b}, the microscopic local pressure tensor for interacting spherical particles without torques is derived using the  steady state probability distribution function $\psi(\mathbf{r},\mathbf{u})=\langle\sum_{i=1}^{N}\delta(\mathbf{r}-\mathbf{r}_i)\delta(\mathbf{u}-\mathbf{u}_i)\rangle$, where $\mathbf{r}=(x,y,z) $
is the 3-dimensional spatial coordinate and unit vector $\mathbf{u}$ is the analogue for orientation.
The diagonal spatial components of this local pressure tensor, $P_{\alpha\alpha}(z)$, consist of an ideal gas contribution, a virial  contribution, and a swim pressure contribution 
\begin{align}\label{SUReq:normal}
 P_{\alpha\alpha}(z)=P^{\rm id}_{\alpha\alpha}(z)+P^{\rm vir}_{\alpha\alpha}(z)+ P^{\rm swim}_{\alpha\alpha}(z).
\end{align}
The ideal component of the pressure reads 
\begin{align}\label{SUReq:ideal}
 P^{\rm id}_{\alpha\alpha}(z)=\rho(z) k_BT_s, 
\end{align}
with $\rho(z)$ the density profile, and the virial and swim contributions due to the particle interactions and self-propulsion are given by 
\begin{align}\label{SUReq:virial3}
 \! \!\!P^{\rm vir, swim}_{\alpha\alpha}(z) = \frac{1}{L^2}\!\!\int\!\! dx\!\!\int\!\! dy P^{\rm vir, swim}_{\alpha\alpha}(\mathbf{r}),
\end{align}
with
\begin{align}\label{SUReq:virial}
 \! \!\!P^{\rm vir}_{\alpha\alpha}(\mathbf{r})\! = -\!\!\int_{{\mathbf r}_{\alpha}''<{\mathbf r}_{\alpha}}\!\!\!\!\!\!\!\!\!\!\!d{\mathbf r}_{\alpha}''\!\int\!\!d{\mathbf r}'\!\!\!~\rho^{(2)}({\mathbf r}''\!\!,{\mathbf r'}) \frac{\partial U(|\mathbf{r}'' \! -\mathbf{r'}|)}{\partial {\mathbf r}''_{\alpha}},
\end{align}
where $\rho^{(2)}({\mathbf r},{\mathbf r'})= \!\!\int\!\! d {\mathbf u}\!\!\int\!\! d {\mathbf{u'}} \psi^{(2)}({\mathbf r},{\mathbf u},\mathbf{r'},\mathbf{u'}) $ is the spatial two-body correlation function with the  full two-body correlation function $ \psi^{(2)}({\mathbf r},{\mathbf u},\mathbf{r'},\mathbf{u'})   \equiv  \langle  \sum_{i=1}^{N}\sum_{j\neq i}^N \delta(\mathbf{r}-\mathbf{r}_i)\delta({\mathbf u}-{{\mathbf u}_i})\delta(\mathbf{r}'-\mathbf{r}_j)\delta({{\mathbf u}}'-{{\mathbf u}_j})   \rangle$. Here, the angular brackets $\langle \dots \rangle$ denote a time  average over steady states. The integral is over the $\alpha$ component of the vector $\mathbf{r}''$ and we define $\mathbf{r}''_\beta=\mathbf{r}_\alpha, \forall \beta\neq\alpha$. 

The local swim pressure contribution is given by:
\begin{align}
P_{\alpha\alpha}^{\mathrm{swim}}(\mathbf{r}) &= \frac{k_BT_sv_0^2}{2D_t D_r} \int\!\! d {\mathbf u}~\psi(\mathbf{r},{\mathbf u}){\mathbf u}_{\alpha}{\mathbf u}_{\alpha}\hfill \nonumber \\
&\hspace{-1cm}- \frac{v_0}{2 D_r}  \!\int  \!\!d {\mathbf {u}} \!\!\int \!\!d {\mathbf r'}\!\!\int \!\!d {\mathbf {u'}} \psi^{(2)}({\mathbf r},{\mathbf {u}},{\mathbf r'},{\mathbf{u'}})  \frac{\partial U(|\mathbf{r} \! -\mathbf{r'}|)}{\partial \mathbf{r}_{\alpha}} \mathbf{u}_{\alpha}  \nonumber \\
& \hspace{-1cm}
 -\frac{k_BT_sv_0}{2D_r} \frac{\partial}{\partial \mathbf{r}_{\alpha}} \! \int \!\! d {\mathbf u}~\psi(\mathbf{r},{\mathbf u}){\mathbf u}_{\alpha}
\end{align}

In our simulations, we measure the density profiles $\rho(z)$ and the normal and transverse components of the pressure tensor profiles, $P_{N}(z)$ and $P_T(z)$  by dividing the system  into small slabs of width $\Delta z=0.1 \sigma$ and area $L^2$ and by measuring the local average quantities in each bin.	
The local density $\rho(z_k)$ in  bin $k$ centered around $z=z_k$ is measured using 
\begin{align}
\rho(z_k) = \frac{\langle n(z_k)\rangle}{\Delta V}, 
\label{eqn:local_idlpress}
\end{align}
with $\langle n(z_k)\rangle$  the average number of particles in bin $k$, and $\Delta V = L^2 \Delta z$ the volume of a  bin. 
Using Ref. \onlinecite{Ikeshoji2003}, we rewrite and measure the virial contribution in bin $k$ as follows 
\begin{align}
P^{\rm vir}_{\alpha\alpha}(z_k)\!=\! \frac{1}{2 \Delta V} \!\left \langle\sum_{i=1}^{n(z_k)}\sum_{j \neq i}^N \frac{{\mathbf r}_{ij,\alpha}}{{r}_{ij}} \frac{d U(r_{ij})}{d r_{ij}}  \!\! \!\!\int_{C_{ij}\in\Delta z_k}  \!\! \!\! \!\! \!\!\! d \mathbf{l}_{\alpha} \right \rangle \!,
\end{align}
with ${r}_{ij} = |\mathbf{r}_{ij}| = |\mathbf{r}_{j}-\mathbf{r}_{i}|$ the center-of-mass distance of particle $i$ and $j$. The variable of integration $\mathbf{l}_{\alpha}$ is along the $\alpha$ component of the integration contour $C_{ij}$ from   ${\mathbf r}_i$ to ${\mathbf r}_j$. The integral  denotes that the virial contribution to the pressure of particle pair $i$ and $j$  is due to the part of $C_{ij}$ that lies inside the respective bin within the coarse-grained Irving-Kirkwood approximation. \cite{Ikeshoji2003}

Finally, the local swim pressure can be measured in each bin $k$ using
	\begin{align}
	{P}_{\alpha\alpha}^{\mathrm{swim}}(z_k) =& \frac{k_BT_sv_0^2}{2D_tD_r\Delta V} \left\langle \sum_{i=1}^{n(z_k)}\mathbf{u}_{i,\alpha}^2 \right\rangle	\nonumber \\
					&- \frac{v_0}{2D_r\Delta V}\left\langle \sum_{i=1}^{n(z_k)}\sum_{j \neq i}^N \frac{{\mathbf r}_{ij,\alpha}}{{r}_{ij}} \frac{d U(r_{ij})}{d r_{ij}} 
					\mathbf{u}_{i,\alpha} \right\rangle	\nonumber \\
					&- \frac{k_BT_sv_0}{2D_r\Delta V}\frac{\partial}{\partial \mathbf{r}_{ij,\alpha}} \left\langle \sum_{i=1}^{n(z_k)} \mathbf{u}_{i,\alpha} \right\rangle
	\label{eqn:jeroen_swimpressure}
	\end{align}
	Note that the last term of Eq.~(\ref{eqn:jeroen_swimpressure}) is a term
	not present in the case of an isotropic bulk phase as discussed previously in Refs. 
	\onlinecite{Solon2015,Falasco2015,Winkler2015}.
	This term is non-zero for systems with finite polarization, defined as $\mathbf{m}(z_k)=\langle \sum_{i=1}^{n(z_k)}\mathbf{u}_{i} \rangle/\Delta V$, for instance at interfaces or surfaces, but  disappears in the homogeneous bulk phase of the fluid.

	\begin{figure}
	\centering		
	\includegraphics[width=\columnwidth]{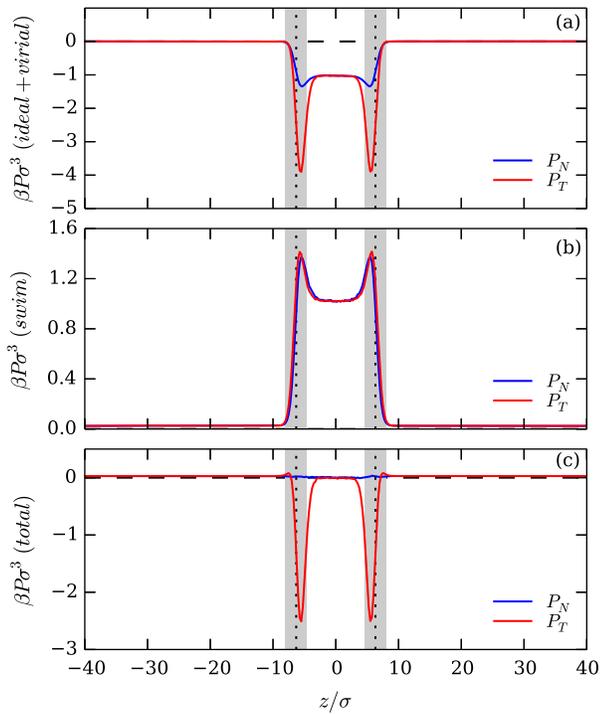}
	\caption{Normal and tangential components of (a) the ideal and virial pressure tensor $P^{\mathrm{id}}_{N,T}(z)+P^{\mathrm{vir}}_{N,T}(z)$, (b) the swim pressure tensor $P^{\mathrm{swim}}_{N,T}(z)$, and 
	(c) the total pressure tensor $P^{\mathrm{tot}}_{N,T}(z)=P^{\mathrm{id}}_{N,T}(z)+P^{\mathrm{vir}}_{N,T}(z)+P^{\mathrm{swim}}_{N,T}(z)$. 
	The normal component of $P^{\mathrm{tot}}_N(z)$ is constant for all $z$ indicating mechanical equilibrium.
	The tangential component shows distinctive peaks at the two interfaces. The simulation parameters are the same as 
	in Fig.~\ref{fig:fig1}
	\label{fig:fig2}}
	\end{figure}

\section{Results}
\label{sec:results}
Using Brownian dynamics simulations, we investigate the interfacial properties of vapour-liquid interfaces in systems of active Lennard-Jones particles for different combinations of self-propulsion speed and rotational diffusion rate, i.e., for varying P\'eclet numbers. 
\subsection{Density and Orientation profiles}
\label{sec:rho_prof}
To start our investigation, we first measure and plot the average density profile $\rho(z)$ to verify  coexistence of vapour and liquid phases in our simulation box. We choose the self-propulsion speed, density, and temperature  such that they correspond to a state point that lies well-inside the two-phase coexistence region as determined  in Ref.~\onlinecite{Prymidis2016}.
A typical configuration of a steady state exhibiting vapour-liquid phase coexistence of  $N=2500$ active Lennard-Jones particles  is shown in Fig.~\ref{fig:fig1}(a), and the corresponding density profile  is presented in Fig.~\ref{fig:fig1}(b). 
We find that the density profiles are similar to
	passive equilibrium profiles and can be well fitted to a hyperbolic tangent function 
	\begin{equation}
	\rho(z) = \frac{1}{2} \left(\rho_l + \rho_v\right) -
				\frac{1}{2} \left(\rho_l - \rho_v\right) \tanh \left[\frac{2(z-z_0)}{D}\right], 
	\label{eqn:tanh_rho}
	\end{equation}
where $\rho_{l}$ and $\rho_{v}$ are the corresponding bulk liquid and vapour coexisting densities,
	$z_0$ is the location of the plane satisfying an equal-area construction and $D$ represents the
	thickness of the interface.
	We fit the above equation to the right and left half of the box ($z>0$ and $z<0$) separately using $z_0$ and $D$ as fitting parameters and obtain the bulk densities $\rho_l$ and $\rho_v$ from the mean of the two fits. We denote the interfacial regions of width $D$ as shaded grey areas in Figs.~\ref{fig:fig1} and \ref{fig:fig2}.
	
In Fig.~\ref{fig:fig1}(c) we plot the local average orientation of particles $\mathbf{u}(z)=\left\langle\mathbf{m}(z)\right\rangle/\left\langle\rho(z)\right\rangle$.  It is evident  that the particles tend to orient themselves along the normal direction at the interfaces, and the peak  of the orientation profile does not coincide with
	the estimated position of the interface $z_0$ (dotted lines).
	On average, the particles tend
	to orient themselves with the direction of self-propulsion towards the less-dense (vapour) phase. This asymmetry in the average orientation is easily explained by assuming a zero net velocity at the interface: particles at the interface that point towards the dense phase have a larger average velocity than particles that point towards the dilute phase due to the net attractive force towards the liquid. Thus, more particles need to point outwards in order to balance the asymmetry in velocities.  It is also important to note that this preferential ordering is only along the normal ($z$) direction.
There is no net orientation along the tangential plane ($xy$) as the system is isotropic in  this plane. We note that in the case of MIPS, where the activity drives the phase separation, the orientations tend to be exactly reverse, with the preferred orientation of particles at the interfaces being towards the denser phase.
	
We also find that at fixed   activity, which for our system translates to fixed self-propulsion speed and rotational diffusion coefficient, the shape  of the orientation profile along the interface as well as the interfacial width $D$  becomes broader upon increasing $T$, or equivalently upon decreasing the strength of attraction  between particles. The broadening of the interface as the system moves towards its ``critical point''  is completely analogous to what is observed in the passive LJ system. \cite{Vrabec2006}
Also, at  fixed temperature $T$, the interfacial region becomes broader as the activity increases. This observation is compatible with  Ref. \onlinecite{Prymidis2016}, which showed that higher attraction strength is needed to induce phase separation upon increasing activity and is also consistent with other older studies. \cite{Schwarz2012,Mognetti2013,Farage2015}

\subsection{Pressure profiles}
\label{sec:pressure_prof}
Subsequently, we  measure the different contributions to the normal and tangential components of the pressure tensor using Eqs. (\ref{eqn:local_idlpress})-(\ref{eqn:jeroen_swimpressure}), 
for our phase-separated systems. Fig.~\ref{fig:fig2} shows typical  profiles of the normal and tangential component of the ideal and virial pressure tensor, the swim pressure tensor, as well as the total pressure tensor. 
Below we discuss the various contributions as well as the total pressure profiles separately.

In passive systems,  mechanical equilibrium requires a constant normal component of the total pressure, which simply consists of an ideal and a virial contribution, in the direction perpendicular to the interfaces. 
However, a net imbalance of the interaction forces along the tangential plane  causes   the tangential component of the pressure to be smaller on average than the normal component along the interfacial region. This inequality of the pressure components at the interface leads to a positive surface tension in the case of equilibrium fluids. \cite{Berry1971,Marchand2011} In the case of our active LJ system, Fig. \ref{fig:fig2}(a) shows that the normal component of the sum of the ideal and the virial pressure is not constant across the system and that the liquid has a smaller bulk pressure than the vapour phase. Thus, mechanical equilibrium is not established simply by considering the virial and the ideal components of the pressure. Moreover, the tangential component is also not equal at the bulk of the two coexisting phases, though it is reassuringly equal to the normal component in the bulk. It is also smaller on average than the normal component of the pressure along the interface, similar  to the passive case. Note that the behavior of the sum of the  ideal and the virial components of the pressure is reversed with respect to their respective profiles in the case of MIPS. \cite{Bialke2015} In that case the ideal and virial component are higher in the dense phase than in the dilute phase.

The swim pressure, as we see in Fig.~\ref{fig:fig2}(b), is also not equal in the two phases for both the normal and tangential components. 
Its magnitude is larger in the liquid phase than the vapour phase where it is essentially zero. Also, both components show peaks along the interfaces. Again, the pressure profile has opposite behavior with respect to the case of MIPS, where the swim pressure is higher in the dilute phase than in the dense phase. \cite{Bialke2015}

In Fig.~\ref{fig:fig2}(c) we show the total pressure, that is the sum of the ideal, the virial and the swim pressure. Reassuringly, the normal component
now becomes constant throughout the system, as is required for mechanical equilibrium.
We wish to emphasize here that the gradient term of the form $\partial_\alpha \mathbf{m}_\alpha$  in the swim pressure, Eq. (\ref{eqn:jeroen_swimpressure}), needs to be included in the total pressure
to obtain a perfectly flat profile for the normal component of the pressure tensor at the interface. This term is obviously zero in the bulk of the system but its magnitude along the interface increases as the activity of the system is increased.
The tangential component of the total pressure is also equal in the two bulks but has negative peaks at the interfaces. This is again similar to the case of equilibrium systems and leads to a positive surface tension, as we will discuss in more detail in Section \ref{sec:surf_tension}. In the case of MIPS the total pressure profiles 
again recover to equal pressures in the bulks upon including the swim pressure but the tangential component has different behavior at the interface than the ones shown in Fig. \ref{fig:fig2}(c).\cite{Bialke2015} The tangential component in that case has positive peaks which translate into a negative vapour-liquid interfacial tension.

  \subsection{Surface Tension}
\label{sec:surf_tension}
In the case of equilibrium fluids,  the surface tension $\gamma$ of an interface that separates two coexisting bulk phases  can be defined in various ways. \cite{Fortini2005} The surface tension can be defined thermodynamically as the difference in grand potential between a phase-separated system with an interface and a homogeneous bulk system, which are both at the same coexisting bulk chemical potential,  divided by the surface area of the interface. Using this definition, the vapour-liquid interfacial tension can be determined in simulations by measuring the grand canonical probability distribution function of observing $N$ particles in a volume $V$ at fixed chemical potential $\mu$ and temperature $T$. This probability distribution function can be measured very accurately using successive umbrella sampling in grand canonical Monte Carlo simulations. \cite{Virnau2004} Using the histogram reweighting technique, one can then determine  the chemical potential corresponding to bulk coexistence using the equal area rule for the vapour and liquid peak.\cite{Virnau2004,Fortini2005} The interfacial tension can be determined from the difference in the maximum of the peaks and the minimum. \cite{Binder1982,Potoff2000,Muller2003}  Alternatively, one can also determine the surface tension by measuring the width of the interface, which is determined by an intrinsic width and a broadening due to capillary wave fluctuations. Using equipartition theorem, one can relate the mean-square fluctuations due to capillary waves to the interfacial tension, and hence the interfacial tension can be determined by measuring the capillary wave broadening. \cite{Sides1999a,lacasse1998capillary,buff1965interfacial,beysens1987thickness,Fortini2005} It is important to note that the method to determine the interfacial tension from the probability distribution is based on  grand canonical Monte Carlo simulations, and relies on a knowledge of the statistical weight corresponding to the grand canonical ensemble. The second method employs the equipartition theorem, which is derived  by assuming a Boltzmann distribution. Finally, the interfacial tension can be defined as the mechanical work required to enlarge the interface. Using the condition of hydrostatic equilibrium, the surface tension can be defined as the integral of the difference of the two pressure tensor components  
	\begin{equation}
	\gamma =\frac{1}{2}\int_{-L/2}^{L/2} \left[P_N(z) - P_T(z)\right]dz, 
	\label{eqn:gamma}
	\end{equation}
where we assume that the system is only inhomogenous in the $z$-direction with the two planar interfaces parallel to the $xy$-plane.  The factor $\frac{1}{2}$ comes  from the presence of two interfaces in a simulation with periodic boundary conditions. For equilibrium fluids, all these definitions for the interfacial tension coincide. 
In the case of non-equilibrium systems such as the active LJ system, the statistical weights of the different ensembles are unknown, which precludes the use of Monte Carlo simulations for determining the interfacial tension from a probability distribution function. We therefore resort to the mechanical definition of the surface tension by employing Eq. (\ref{eqn:gamma}). In addition, we  measure the interfacial width  in Brownian dynamics simulations, and naively assume the equipartition theorem to hold even though it is based on a statistical ensemble average. We present and discuss our results below using the mechanical definition and later in Section~\ref{sec:capillary} for the application of capillary wave theory.

	\begin{figure}
	\centering
	\includegraphics[width=\columnwidth]{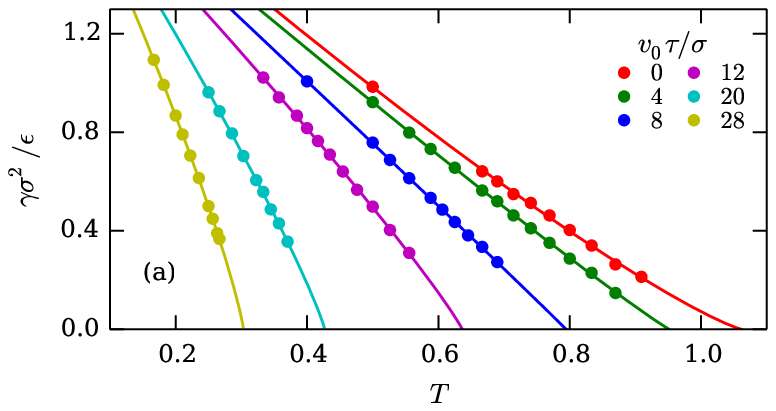}
	\includegraphics[width=\columnwidth]{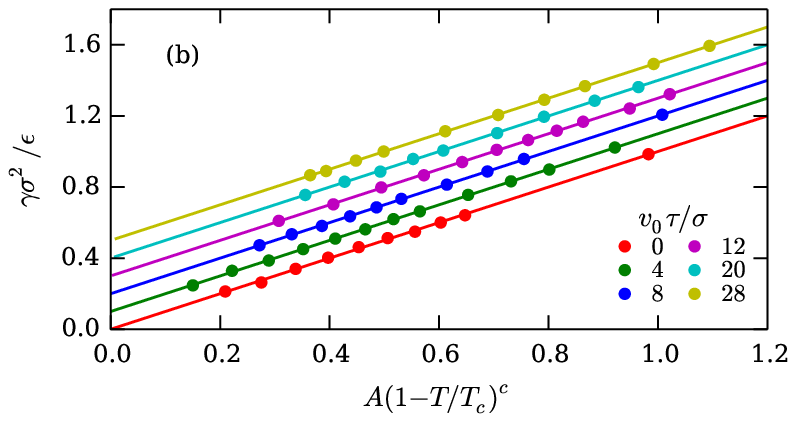}
	\begin{tabular}[b]{c}
			\includegraphics[width=0.47\columnwidth]{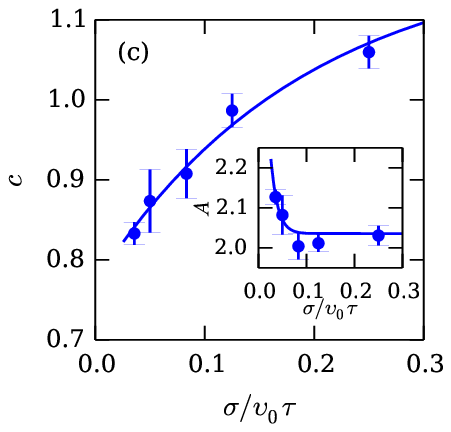}
	\end{tabular}
	\begin{tabular}[b]{c}
			\includegraphics[width=0.47\columnwidth]{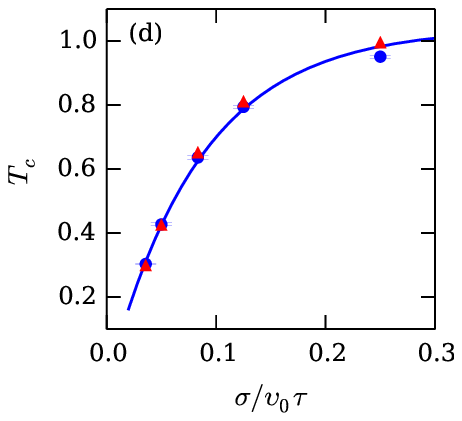}
	\end{tabular}
	\caption{vapour-liquid interfacial tension $\gamma\sigma^2/\epsilon$ as a function of (a) temperature $T=k_BT_s/\epsilon$ and (b) scaled temperature, for an active Lennard-Jones system 
	(circles) with a rotational diffusion rate  $D_r\tau=20$ for varying self-propulsion
	speeds $v_0 \tau/\sigma$ as obtained from Eq.~(\ref{eqn:gamma}) and corresponding fits using Eq.~(\ref{eqn:fit_gam}). Results in (b) are offset for clarity.
	(c) Fit parameters $c$ and $A$ (inset) as a function of  the inverse self-propulsion speed $\sigma/(v_0\tau)$ with errorbars in the estimate of these parameters,
	and the corresponding fit using Eq.~(\ref{eqn:fit_v0}). 
	(d) The scaling of the estimated $T_c$  as obtained from the scaling of the surface tension $\gamma$ 
	(circles) and the
	values obtained from the scaling of the coexistence densities from Ref.~\onlinecite{Prymidis2016}
	(triangles)
	along with the corresponding fits using  Eq.~(\ref{eqn:fit_v0}).
	\label{fig:fig3}}
	\end{figure}

	\begin{figure}
	\centering
	\includegraphics[width=\columnwidth]{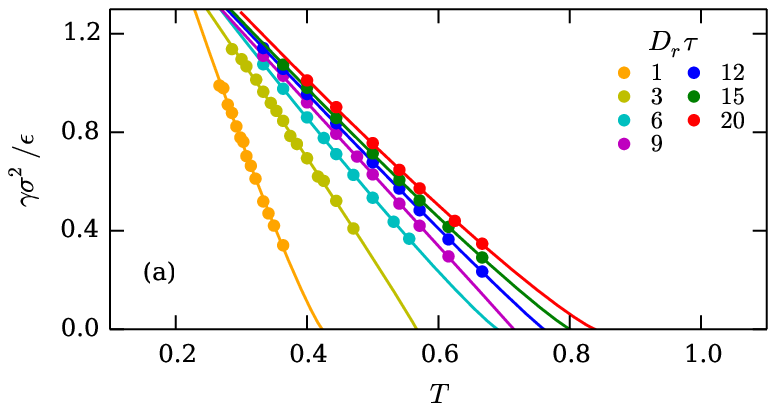}	
	\includegraphics[width=\columnwidth]{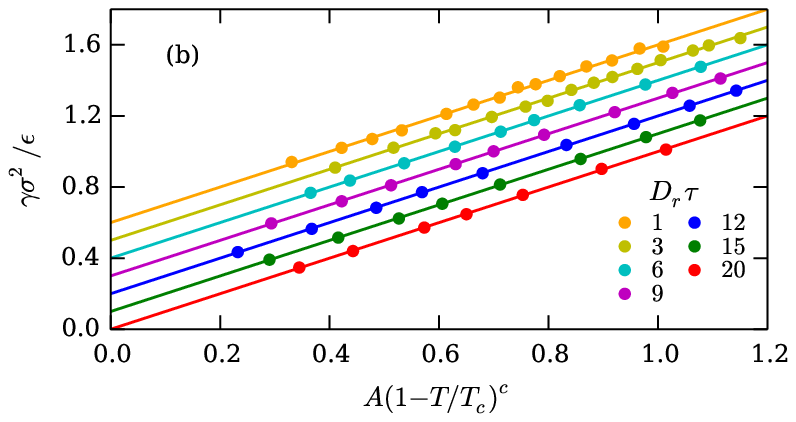}
	\begin{tabular}[b]{c}
			\includegraphics[width=0.47\columnwidth]{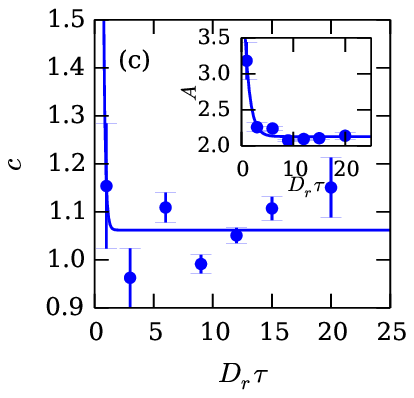}
	\end{tabular}
	\begin{tabular}[b]{c}
			\includegraphics[width=0.47\columnwidth]{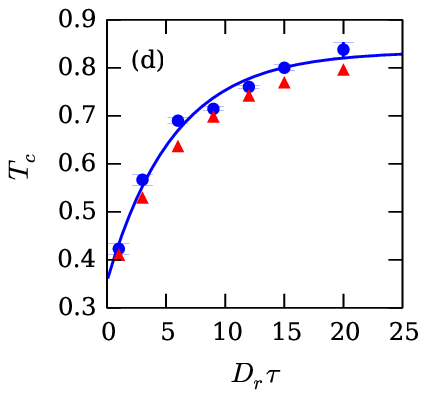}
	\end{tabular}
	\caption{vapour-liquid interfacial tension $\gamma\sigma^2/\epsilon$ as a function of (a) temperature $T=k_BT_s/\epsilon$ and (b) scaled temperature, for an active Lennard-Jones system (circles) with a self-propulsion
	speed $v_0\tau\sigma^{-1}=8$ and varying rotational diffusion rates  $D_r\tau$ as obtained from Eq.~(\ref{eqn:gamma}) and corresponding fits using Eq.~(\ref{eqn:fit_gam}). Results in (b) are offset 
	for clarity.
	(c) Fit parameters $c$ and $A$ (inset) as a function of  the rotational diffusion rates $D_r\tau$ with errorbars,
	and the corresponding fit using Eq.~(\ref{eqn:fit_dr}). 
	(d) The scaling of the estimated $T_c$  as obtained from the scaling of the surface tension $\gamma$ 
	(circles) and the
	values obtained from the scaling of the coexistence densities from Ref.~\onlinecite{Prymidis2016}
	(triangles)
	along with the corresponding fits using  Eq.~(\ref{eqn:fit_v0}).
	\label{fig:fig4}}
	\end{figure}
	
Following Ref.~\onlinecite{Bialke2015},  we determine the surface  tension using the mechanical route  (Eq. (\ref{eqn:gamma})), where we also include the contribution from the swim pressure in the total pressure in order to satisfy the hydrostatic equilibrium condition. Note that the gradient term of the form $\partial_\alpha \mathbf{m}_\alpha$  in the swim pressure (Eq. (\ref{eqn:jeroen_swimpressure})), which is essential in order to obtain a flat profile of the normal pressure component across the interface, does not contribute to the surface tension. Using   Eq. (\ref{eqn:gamma}) and the total pressure profiles as exemplarily shown in Fig. \ref{fig:fig2}(c) we determine the surface tension $\gamma$ for a wide range of parameters of the active system following two paths that drive the system out of equilibrium. To this end,  we either increase the self-propulsion speed at a fixed  rotational
diffusion rate ($D_r\tau=20$) or we decrease the rotational diffusion coefficient at a fixed self-propulsion speed ($v_0\tau\sigma^{-1}=8$).
For the first path, we choose a high value for the rotational diffusion coefficient in order to minimize the regime of  percolating states in the state diagram. \cite{Prymidis2015b}
Note that the equilibrium limits of  these two paths  are not equivalent as the $D_r\rightarrow\infty$  	limit does not coincide with the $v_0\rightarrow 0$ limit.
The second limit	corresponds to the equilibrium LJ system with temperature $k_BT_s/\epsilon$  while the first limit corresponds to a passive system with a higher effective temperature. The systems we examine have a P\'eclet number in the range $0-8$, where the P\'eclet number is defined as ${\textrm Pe}=v_0/D_r\sigma$, so that we probe the equilibrium limit as well as systems where self-propulsion plays a much more important role in the dynamics than translational diffusion. However,  in all cases we are well below  the onset of MIPS\cite{Stenhammar2014}, i.e, $\textrm{Pe}\sim 50$.
	
We plot the surface tension $\gamma\sigma^2/\epsilon$ as a function of temperature $T=k_BT_s/\epsilon$  in Figs. \ref{fig:fig3}(a) and \ref{fig:fig4}(a) for constant rotational diffusion coefficient and constant speed of self-propulsion, respectively.  Note that we always measure a positive surface tension, contrary to the case of MIPS \cite{Bialke2015} and the magnitude of the surface tension is in the same range  ($\gamma\sigma^2/\epsilon \sim 1$) as in the equilibrium system.\cite{Vrabec2006} We also find that the surface tension decreases upon increasing the temperature towards the critical temperature $T_c$ as the density difference between the coexisting phases decreases, which is similar to  equilibrium systems for which the surface tension vanishes at the critical point. 

Next, we examine  the scaling of the surface tension $\gamma$ with  temperature $T$ as the system departs from the equilibrium regime by increasing the  activity of the Lennard-Jones particles. In the case of equilibrium systems,  $\gamma$ scales with temperature as 
	\begin{equation}
	\gamma\sigma^2/\epsilon = A(1-T/T_c)^{c},
	\label{eqn:fit_gam}
	\end{equation}	
where $A$ denotes a dimensionless constant, $T_c$  the critical temperature, and $c$ a critical exponent. In the case of equilibrium systems, $c=2\nu$ with $\nu=0.63$  the critical exponent of the bulk correlation length of the system. \cite{Watanabe2012}  Here, we examine whether the surface tension for our active system follows a  scaling with temperature similar to Eq. (\ref{eqn:fit_gam}) and treat $A,$ $T_c$ and the exponent $c$ as fit parameters. 
	
In Figs.~\ref{fig:fig3}(b) and \ref{fig:fig4}(b), we plot the resulting fits which are offset for clarity.
The same fits are shown as solid lines in Figs.~\ref{fig:fig3}(a) and \ref{fig:fig4}(a).  We find that they fit well to the measured data in the examined parameter space. We thus observe that the scaling of the surface tension with temperature can be captured by Eq.~(\ref{eqn:fit_gam}) even for the active systems considered here. Note that these fits also give us an estimate for the critical temperature of the system in the limit $\gamma=0$ for different degrees of activity.

	\begin{table}
	\caption{Fitting parameters of Eq.~(\ref{eqn:fit_v0}) for an active Lennard-Jones system  with a rotational diffusion rate $D_r\tau=20$ and
	 varying self-propulsion speeds $v_0\tau\sigma^{-1}$.}
	\centering
	\begin{tabularx}{\columnwidth}{c||Y|Y|Y|Y}
			 & $A$ 	& $c$ & $T_c$ & $T_c\footnotemark[1]$\\
	\hline
	$a_1$ 	& 1.159 & -0.410 & -1.113  & -0.184 \\
	$a_2$ 	& 70.01 &  5.257 &  11.810 & 12.33 \\
	$a_3$ 	& 2.035 &  1.181 &  1.041  & 1.066 \\
	\hline
	\end{tabularx}
	\label{tab:crit_exp_v0}

	\caption{Fitting parameters of Eq.~(\ref{eqn:fit_dr}) for an active Lennard-Jones system  with a  self-propulsion speeds $v_0\tau\sigma^{-1}=8$ and varying rotational diffusion rates 
	$D_r\tau$.}	
	\centering
	\begin{tabularx}{\columnwidth}{c||Y|Y|Y|Y}
			 & $A$ 	& $c$ & $T_c$ & $T_c\footnotemark[1]$ \\
	\hline
	$b_1$ 	& 2.840 & 109.99 & -0.478 & -0.470 \\
	$b_2$ 	& 0.993 &  7.095 &  0.178 & 0.156 \\
	$b_3$ 	& 2.128 &  1.062 &  0.834 & 0.818 \\
	\hline
	\end{tabularx}
	\label{tab:crit_exp_dr}	
	
	\footnotetext[1]{values from Ref.~\onlinecite{Prymidis2016}}
	\end{table}

We now examine the scaling of the fit parameters  $A,$ $c$ and $T_c$  upon increasing the activity. The results for
the parameters $A$ and $c$ are plotted in Figs. \ref{fig:fig3}(c) and \ref{fig:fig4}(c).  We find that driving the system further away from equilibrium by increasing the self-propulsion speed at fixed rotational diffusion coefficient, the value of the exponent $c$ decreases, while the parameter $A$ increases (Fig.~\ref{fig:fig3}(c)). The exponent $c$ moves away from its equilibrium value $c=1.21-1.26$ \cite{Vrabec2006,Anisimov1991critical} to values less than unity. 
We find a similar scaling in the case where the system is driven out of equilibrium by decreasing the rotational diffusion coefficient at fixed self-propulsion speed, i.e., $c$ decreases, while $A$ increases. However, the exponent $c$ appears to increase again  for very low values of the rotational diffusion coefficient as shown in Fig. \ref{fig:fig4}(c). Unfortunately, large errorbars in the fits for this regime prevent us from making any definite conclusions on the  dependence of the exponent $c$ on the activity of the system for high P\'eclet numbers. 

Furthermore, the scaling of the critical temperature with the self-propulsion force is in accordance with the findings of Ref. \onlinecite{Prymidis2016} showing that $T_c$ decreases upon increasing activity. In Fig.~\ref{fig:fig3}(d) and \ref{fig:fig4}(d) we plot the cases of varying self-propulsion speed and varying rotational diffusion coefficient respectively, both demonstrating the trend.
Lastly, we also compare the critical temperature as determined from the scaling of the order parameter $\Delta\rho=\rho_l-\rho_v$ from Ref. \onlinecite{Prymidis2016} in Figs.~\ref{fig:fig3}(d) and ~\ref{fig:fig4}(d). We find that the two values of the critical temperature as evaluated from the two different routes, i.e, via the scaling of the order parameter and via the scaling of the surface tension with temperature, are very close to each other
in the case of varying self-propulsion speed but the agreement is not as good in the case of varying rotational diffusion at constant self-propulsion  speed. Nonetheless, the values from the two routes are still close to one another and follow a similar scaling.

Additionally, we show empirical fits for the dependence of the parameters $A$, $c$ and $T_c$  on the self-propulsion speed $v_0$ and the rotational diffusion coefficient  $D_r$, respectively. All three parameters are fitted by simple exponential scalings, namely 
	\begin{align}
	A(v_0),c(v_0), T_c(v_0) &= a_1e^{-a_2\sigma/v_0\tau}+a_3	\label{eqn:fit_v0} \\
	A(D_r), c(D_r),T_c(D_r) &= b_1e^{-b_2 D_r\tau}+b_3		\label{eqn:fit_dr},
	\end{align}
where $a_1,a_2,a_3$ and $b_1,b_2,b_3$ are again fit parameters. These fits  capture the scaling of the exponent $c$ and the parameter $A$ (Figs.~\ref{fig:fig3}(c) and  \ref{fig:fig4}(c)) for varying self-propulsion speeds $v_0$ (Eq. \ref{eqn:fit_v0}) and rotational diffusion coefficients $D_r$ (Eq. \ref{eqn:fit_dr}). The fit for $c$ obviously fails for  varying  rotational diffusion coefficient (Fig.~\ref{fig:fig4}(c)), but we still present it for consistency.
The scaling of the critical temperature $T_c$ is shown in Fig.~\ref{fig:fig3}(d) and ~\ref{fig:fig4}(d) along with the values from Ref.~\onlinecite{Prymidis2016}.
The fit parameters $a_1,a_2,a_3$ and $b_1,b_2,b_3$ providing the scaling of $A,$  and $c$ as well as the two values for $T_c$ are listed in Tables~\ref{tab:crit_exp_v0} and \ref{tab:crit_exp_dr} for varying self-propulsion speeds and rotational diffusion coefficients, respectively.

Before closing this section, it is important to remark that we use Eq. (\ref{eqn:fit_gam}) merely as a fit	to our results, and that we do not identify the associated fit parameters with the critical point and critical exponents of the current system.  In fact, we have not yet demonstrated the existence of a critical point for active LJ systems as we are unable to obtain reliable data on the vapour-liquid phase coexistence in the critical regime due to the small system sizes that we used in our simulations. Nonetheless, it is instructive to compare  the equilibrium limit of our measurements to their known equilibrium values, with the equilibrium limit corresponding to the limits  $v_0 \tau/\sigma\rightarrow 0$ and $D_r\tau\rightarrow \infty$   for the results presented in Tables~\ref{tab:crit_exp_v0} and \ref{tab:crit_exp_dr}, respectively.    We find that our estimates for the critical point and exponents are rough, yet reasonable. Specifically, we estimate the equilibrium critical point at $T_c = 1.041$, while recent finite size scaling studies report $T_c = 1.187$.\cite{shi2001histogram} 
Furthermore, we find  the exponent $c=1.181$ and $1.062$, while literature reads $c=1.21-1.26$. \cite{ Anisimov1991critical,Vrabec2006}

\subsection{Interface fluctuations and Stiffness}
\label{sec:capillary}
	In this section we study the scaling of the interfacial width as a function of the area of the interface, which allows us to measure the stiffness of the interface. Subsequently we attempt to connect the estimated value for the stiffness to the value of the surface tension obtained in Section \ref{sec:surf_tension}.
	
For equilibrium systems, capillary wave theory provides a connection between the fluctuations of an interface and its stiffness coefficient or interfacial tension. \cite{Sides1999a,lacasse1998capillary,buff1965interfacial,beysens1987thickness} Capillary wave theory \cite{rowlinson2002molecular} describes the broadening of an intrinsic interface of width $w_0$ due to thermal fluctuations. The capillary wave broadening depends primarily on the interfacial tension and the area of the interface, and can be calculated by using equipartition theorem and summing over the mean-square fluctuations of all allowed excitation modes of the interface. We refer the reader to Refs. \onlinecite{Sides1999a,lacasse1998capillary} for more details, and present here only the result. 
According to  capillary wave theory \cite{Sides1999a,lacasse1998capillary} the total interfacial width $w$ can be written as the sum of an intrinsic part $w_0$ and a contribution due to capillary wave fluctuations
\begin{equation}
w^2 = w_0^2+\frac{1}{\kappa} \ln\left( \frac{L}{\xi}\right),
\label{eqn:capillary}
\end{equation}
where $\xi$ is the bulk correlation length and $\kappa$ is the stiffness coefficient, which parametrizes the energy penalty for deformations of the interface with dimensions $L\times L$. Eq. (\ref{eqn:capillary}) implies that the width of an interface is determined by an intrinsic contribution $w_0$ that depends only on intensive variables  and a term that explicitly depends on the area of the interface.  For equilibrium systems, the stiffness coefficient $\kappa$ of an interface that separates two coexisting fluids is related to the surface tension \textit{via} the simple relation $\gamma=k_BT_s\kappa$.
	
First, we test the applicability of Eq.~(\ref{eqn:capillary}) to our out-of-equilibrium system. To this end, we perform simulations with phase-separated systems of different sizes such that the area of the planar interface is set at  $L^2$, $2L^2$, $4L^2$, $9L^2$ and $16L^2$, with $L=14.7\sigma$. The smaller area corresponds to the system of 2500 particles that we studied in Sections \ref{sec:rho_prof}-\ref{sec:surf_tension}, while the bigger system has approximately 40000 particles. As we increase the system size we find that the value of the surface tension reassuringly does not change, indicating that the results presented in the previous Section~\ref{sec:surf_tension} are free from large finite size effects. 
In order to measure the width of the interface we first measure the density profile of the various systems. We find that, as the system size is increased, Eq.~(\ref{eqn:tanh_rho}) does not describe our simulation data as accurately as the error function fit
	\begin{equation}
	\rho(z) = \frac{1}{2} \left(\rho_l + \rho_v\right) - 
				\frac{1}{2} \left(\rho_l - \rho_v\right) {\rm erf} \left[\frac{\sqrt{\pi}(z-z_0)}{w}\right],
	\label{eqn:erf_rho}
	\end{equation}
	where the various parameters have the same interpretation as in Eq.~(\ref{eqn:tanh_rho}). This observation has also  been made for passive LJ systems.\cite{Ismail2006} Thus, in this section we use Eq.~(\ref{eqn:erf_rho}) in order to fit the density profiles $\rho(z)$ and estimate the width of the interface $w$ for different systems.

\begin{figure}
	\centering
		\includegraphics[width=\columnwidth]{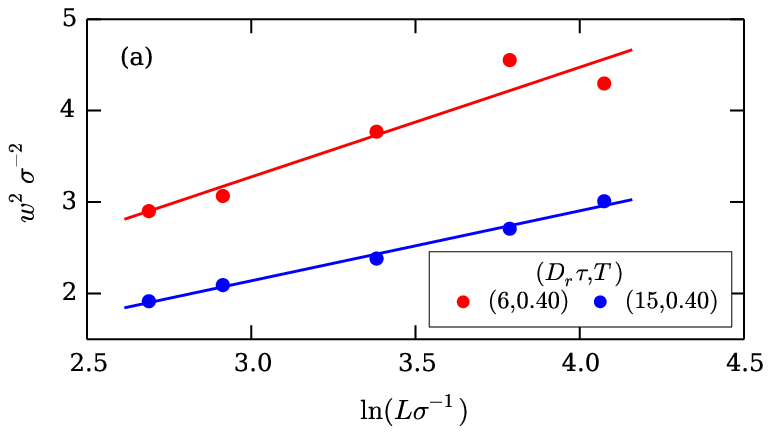}
		\includegraphics[width=\columnwidth]{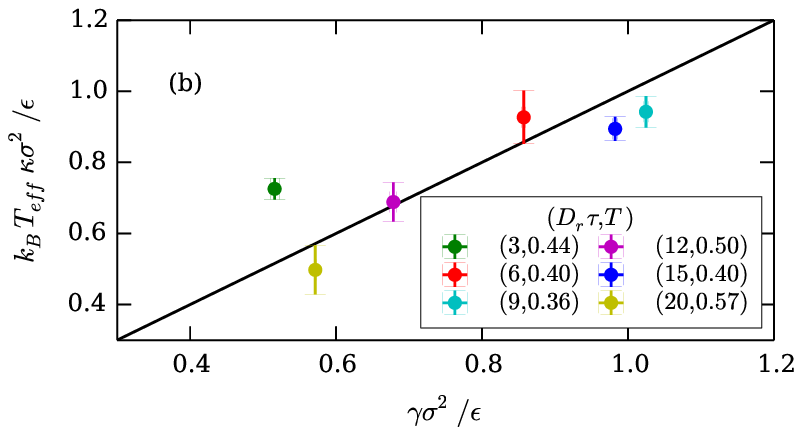}
		\caption{(a) Interfacial width $w$  as a function of the  length $L$ of the interface.  Data  points correspond to simulation results and the continuous lines are fits of Eq.~\ref{eqn:capillary}. 
	(b) The surface tension values measured  from the stiffness coefficient $\kappa$   \textit{via} Eq.~(\ref{eqn:gamkapv0}) versus the interfacial tension as obtained  \textit{via} the pressure tensor route (Eq.~(\ref{eqn:gamma})) for a range of different temperatures $T$ and rotational diffusion coefficients $D_r\tau$ as labeled.  The black line shows the expected scaling $\gamma=k_BT_{\text{eff}}\kappa$. The speed of the self-propulsion is  $v_0\tau\sigma^{-1}=8$ for all systems shown in this figure. Errors bars denote the error in the calculation of  $\kappa$ from the fit of the width Eq. (\ref{eqn:capillary}).
	\label{fig:fig5}}
	\end{figure}

	We perform  simulations for systems of different sizes for various combinations of the self-propulsion speed $v_0$, the rotational diffusion coefficient $D_r$ and temperature $T$. Interestingly, we find that the width of the interface squared indeed scales linearly with the logarithm of the interfacial area, as Eq.~(\ref{eqn:capillary}) prescribes.  In Fig.~\ref{fig:fig5}(a) we plot typical results for two sets of simulation parameters as well as the fit using Eq.~(\ref{eqn:capillary}). These fits allow us to extract the stiffness coefficient $\kappa$.  Note that an equilibrium-like scaling of the width of the interface as a function of the interfacial area has previously been observed  in the case of MIPS in a two-dimensional system.\cite{Bialke2015}

	Next, we compare the value of the stiffness coefficient $\kappa$ as extracted from the scaling  of the width of the interface with the area of the interface to the values of the surface tension $\gamma$ obtained by integrating the pressure tensor profiles (Eq.~(\ref{eqn:gamma})) of the same system. The values of the two quantities have been acquired \textit{via} independent measurements. Remarkably, we find that the two values can be related for all systems  studied  \textit{via} the simple relation
	\begin{align}
	 \gamma &= \left( k_BT_s+  \frac{\eta v_0^2}{6D_r}  \right)\kappa 	\nonumber\\
			&= k_BT_{\text{eff}}\kappa,
	\label{eqn:gamkapv0}
	\end{align}

	where we have defined an effective temperature $T_{\text{eff}}=T_s + \eta v_0^2/6k_BD_r $. Note that this quantity has  already been discussed in literature as a means to connect active systems to their equilibrium counterparts; \cite{Takatori2015,Ginot2015a,Speck2016} ideal passive particles with temperature $T_{\text{eff}}$ share on average the same translational diffusion rates as active Brownian particles with temperature $T_s$, self-propulsion force $v_0$ and rotational diffusion $D_r$. In Fig.~\ref{fig:fig5}(b) we show the comparison between the scaled stiffness coefficient $ k_BT_{\text{eff}}\kappa$ as obtained from the scaling of the interfacial width and the surface tension $\gamma$ measured from the pressure tensor profiles for various system parameters. 
	The figure confirms the applicability of  Eq.~(\ref{eqn:gamkapv0}) to our system, which we have further verified  for various other system parameters (not shown here) and whose effective temperature $T_{\text{eff}}/T_s$ ranges from 1 up to 100.
	As a final remark, note that Bialk\'e \textit{et al}. argue  that a similar relation to Eq.~(\ref{eqn:gamkapv0}) holds also in the case of MIPS,  \cite{Bialke2015} where $\gamma=-\kappa\eta v^2_0/D_r$ in two dimensions. However, an extra minus sign has to be included in this relation since the stiffness coefficient is positive while the 
	surface tension is  negative.


\section{Conclusions}
\label{sec:conclusions}
In conclusion, we performed Brownian dynamics simulations of a three-dimensional system of self-propelled particles interacting with  Lennard-Jones interactions 
at state points that are well-inside the vapour-liquid phase coexistence region. We examine systems with a P\'eclet number $0\leq v_0/D_r\sigma \leq 8$, so that we probe the equilibrium limit as well as systems that are out-of-equilibrium. However, in all cases the phase separation is driven by the cohesive energy of the particles. 

We studied the phase coexistence of a vapour and a liquid phase in an elongated   simulation box and investigated the  properties of the system and the interface. 
By employing a local expression of the pressure tensor for active systems, we measured the normal and tangential components of the pressure tensor in the direction perpendicular to the interface. We verified mechanical equilibrium of the two coexisting phases by measuring a constant normal component of the pressure tensor in the direction perpendicular to the interface. 	 
The tangential component showed negative peaks  at the interface, behaviour reminiscent of equilibrium systems and indicative of a positive non-equilibrium interfacial  tension of the interface as measured by integrating the difference of the normal and tangential component of the pressure tensor.  

We calculated the non-equilibrium surface tension  for different
combinations of self-propulsion speed and rotational diffusion
rate, and demonstrated that the trends of the surface tension can be fitted by simple power laws similar to  equilibrium systems. These scaling laws enabled us to obtain an estimate for the critical temperature of the system as well. Interestingly, the resulting critical temperature of the active system was in close agreement with the values of the critical temperature obtained from the scaling of the order parameter. \cite{Prymidis2016} This agreement  suggests on one hand that the definitions of pressure and surface tension that were used constitute useful tools for the study of the physics of the phase transition and  on the other hand  hints to  a deeper but not yet understood connection between the physics of the passive and the active system. 
	
Furthermore, we calculated the stiffness coefficient of the interface and found a simple equation that relates it to the surface tension. This relation had the same form as in equilibrium systems,  assuming an effective temperature of the interfacial fluctuations. Our results show many similarities  between  bulk and interfacial properties of  active and passive Lennard-Jones systems for state points in the vapour-liquid coexistence region. We hope that, by bringing these similarities into light, we  inspire and assist theoretical work in the direction of building a statistical physics of active matter and its associated  phase transitions.


\section{Acknowledgements}
	S.P. and M.D. acknowledge the funding from the Industrial Partnership Program ``Computational  Sciences  for  Energy  Research'' (grant no.14CSER020) of the Foundation for Fundamental Research on Matter (FOM), which is part of the Netherlands Organization for Scientific Research (NWO). This research programme is co-financed by Shell Global Solutions International B.V.
	V.P. and L.F. acknowledge funding from the Dutch Sector Plan Physics and Chemistry, and L.F.
	acknowledges financial support from the Netherlands Organization for Scientific Research (NWO-VENI
	grant no. 680.47.432). We would also like to thank Berend v. d. Meer for careful reading of the manuscript and Robert Evans for useful discussions.

\bibliography{locallibrary}

\end{document}